\documentclass[grl]{agu2001}
\usepackage{epsf}
\usepackage{rotating}

\authorrunninghead{KR\"UGER ET AL.}
\titlerunninghead{JOVIAN DUST STREAMS: PROBES OF IO PLASMA TORUS}

\journalid{} \articleid{}{} \paperid{}
\cpright{AGU}{2002}

\authoraddr{H. Kr\"uger, MPI Kernphysik, Postfach 103980, 69029 
Heidelberg, Germany
(Harald.Krueger@mpi-hd.mpg.de)}

\authoraddr{M. Hor{\'a}nyi, LASP, 
U. of Colorado, Boulder, CO 80309, USA
(Mihaly.Horanyi@lasp.colorado.edu)}

\authoraddr{E. Gr\"un, MPI Kernphysik, Postfach 103980, 69029
Heidelberg, Germany
(Eberhard.Gruen@mpi-hd.mpg.de)}

\begin{document}

\title{Jovian dust streams: Probes of the Io plasma torus}

\author{Harald Kr\"uger}

\affil{Max-Planck-Institut f\"ur Kernphysik, Postfach 103980, 69029 
Heidelberg, Germany}

\author{Mih{\'a}{\l}y Hor{\'a}nyi}

\affil{Laboratory for Atmospheric and Space Physics, and \\ Department of Physics,
University of Colorado, Boulder, CO 80309, USA}

\author{Eberhard Gr\"un}

\affil{Max-Planck-Institut f\"ur Kernphysik, Postfach 103980, 69029
Heidelberg, Germany, and \\
Hawaii Institute of Geophysics and Planetology,
Univ. of Hawaii,
1680 East West Road,
Honolulu, HI 96822, USA}

\begin{abstract}

Jupiter was discovered to be a source of high speed dust particles 
by the Ulysses spacecraft in 1992. These dust particles originate 
from the volcanic plumes on Io. They collect electrostatic charges 
from the plasma  environment, gain energy from the co-rotating electric 
field of the  magnetosphere, and leave Jupiter with escape speeds over 
$\rm 200\,km\,s^{-1}$.
The dust streams were also observed by the Galileo and Cassini spacecraft.
While Ulysses and Cassini 
only had a single  encounter with Jupiter,  Galileo has 
continuously monitored
the dust streams in the Jovian magnetosphere 
since 1996. The observed dust fluxes exhibit large 
orbit-to-orbit variability due to both systematic and stochastic changes.
By combining the entire
data set,  the variability due to stochatic processes can be approximately 
removed 
and a strong variation with Jovian
local time is found. 
This result is consistent with theoretical expectations and 
confirms that the majority of the Jovian dust stream particles
originate from Io rather than other potential sources.

\end{abstract}

\begin{article}

\section{Introduction}

%
When active volcanism was discovered on Io with the Voyager spacecraft in 
1979 it was soon speculated whether tiny dust grains entrained in Io's plumes
might be ejected into circumjovian space by electromagnetic forces 
[\markcite{{\it Johnson et al.}, 1980, {\it Morfill et al.}, 1980}].
No observational proof for this mechanism, however, was available at that time.

%
In 1992 streams of dust particles originating from the Jovian system
were discovered with the dust detector on-board the Ulysses spacecraft
[\markcite{{\it Gr\"un et al.}, 1993}].
The dust streams consist of approximately 10 nm sized, electrically 
charged, dust grains 
[\markcite{{\it Zook et al.}, 1996}], and they were later shown to originate
from Io [\markcite{{\it Graps et al.}, 2000}].
These tiny grains
gain energy from the corotational electric field of the magnetosphere and
escape from Jupiter with speeds exceeding $\rm 200\,km\,s^{-1}$. 

The stream particles
were also detected by the Galileo and Cassini spacecraft. Since
1996 Galileo made 33 orbits about Jupiter and provided a long record
about the temporal and spatial variability of the ejected dust fluxes.  
These observations show  a large orbit-to-orbit variation 
due to both systematic and stochastic changes.
Systematic effects include Io's orbital motion, changes in
the geometry of Galileo's orbit and in the magnetic field configuration
due to the rotation of the planet.  Stochastic
variations include fluctuations of Io's volcanic activity and the deformation 
of the outer magnetosphere in
response to the variable solar wind conditions.  Using the long record of
Galileo's dust measurements, the variability due to stochastic variations
can be approximately removed by `averaging'.  

Time series
analysis of the  data  collected  through the first two years of
Galileo in orbit about Jupiter revealed that the intensity of the dust fluxes
is modulated by both  Io's orbital period and
Jupiter's rotational period, even though these could not be easily
recognized in single orbits [\markcite{{\it Graps et al.}, 2000}].
In this paper we examine the average spatial
dependence of the dust emission pattern using the entire Galileo data set.
Based on our earlier modeling 
[\markcite{{\it Hor{\'a}nyi et al.}, 1997}], we expect a strong asymmetry 
in the dust emission pattern due to the  asymmetry of the
 parameters of the Io plasma torus.

The plasma torus shows a dawn-dusk brightness asymmetry that
is most likely the result of a large scale cross tail electric field
related to the global flow pattern in the plasma tail of Jupiter
[\markcite{{\it Barbosa and Kivelson}, 1983, {\it Ip and Goertz}, 1983}].
The dawn-dusk electric field shifts the
entire plasma torus by $0.15\,\rm R_J$ (Jupiter radius $\rm R_J=
71,492\,\rm km$) in the dawn direction, resulting in a higher plasma
temperature on the dusk side. The charge of a dust particle escaping the
plasma torus is a function of Io's position in both magnetic and inertial
coordinates, reflecting the plasma density and temperature variations
along Io's orbit. The cross-tail electric field is small compared to the
co-rotational electric field and its direct effect on the dynamics of the
dust particles is small. However, the displacement of the plasma torus
leads to a strong dawn-to-dusk asymmetry in the plasma 
parameters that influences the escape of the dust particles. 

The charge carried by a dust grain will not stay constant as
it travels through the Io plasma torus. Its charge is governed by 
a balance between electron and ion currents
and secondary and photoelectron emission currents which are all functions
of the plasma parameters, material properties, size, velocity and the
instantaneous charge of a dust particle [\markcite{{\it Whipple}, 1981}].
For sufficiently high electron temperatures, 
secondary electron 
production becomes important so that incoming electrons
can generate more than one outgoing secondary electron. Hence, the 
net current can turn positive [\markcite{{\it Hor{\'a}nyi et al.}, 1997}]. 
In general, the charges of
the grains are more positive on the dusk side of the torus where the
electron temperature is higher.

Numerical simulations 
[\markcite{{\it Hor{\'a}nyi et al.}, 1997}] show that dust
particles starting on the dawn side of Jupiter remain captured by the Io
torus for some time whereas particles released at the dusk side have a
higher chance to escape. This leads to a strong variation of the expected
dust flux with Jovian local time 
even if the dust ejection rate of Io
remains constant.

The Galileo dust measurements since 1996  
provide a unique
opportunity for long-term in-situ studies of the Jovian dust environment.  
In this paper we analyze this six-year dust
data set for systematic variations with Jovian local time and compare our 
observations to our current models.

\section{Galileo dust measurements at Jupiter}

Galileo is a dual-spinning spacecraft with an antenna that
points anti-parallel to the positive spin axis (PSA). During most 
of Galileo's orbital tour about Jupiter the antenna pointed towards Earth.
The Galileo Dust Detector System (DDS), like its twin on-board 
Ulysses, is a multi-coincidence impact ionization 
detector 
[\markcite{{\it Gr\"un et al.} 1992a}]
which 
measures submicrometer- and micrometer-sized dust particles.
The DDS is mounted on the spinning section of Galileo 
and the sensor axis is offset by an angle of 60\deg\ 
from the PSA. DDS has a 140\deg\ wide field of view. 
Thus, during one spin revolution of the spacecraft, the
detector scans the entire anti-Earth hemisphere, whereas particles
approaching from the Earth-ward hemisphere remain undetectable.

The dust detector and the procedure for identification of Jupiter stream 
particles in the Galileo dust data set have been
frequently discussed in the literature and are not repeated here
[\markcite{{\it Gr\"un et al.}, 1995, 1998, {\it Kr\"uger et al.,} 1999a, 2001}]. 
In this paper we will use class~3 (our highest
quality class) and de-noised class~2 dust data 
[\markcite{{\it Kr\"uger et al.,} 1999b}].

The Galileo trajectory about Jupiter is shown in a Jupiter-centered 
coordinate system in Fig. 1. 
The spacecraft orbit 
is aligned along the planet's equatorial plane to within a few degrees.
Due to the planet's orbital motion about the Sun, the line of
apsides shifts by 360\deg\
within one Jovian year (11.9 Earth years). Since 1995, when Galileo
was injected into a bound orbit about Jupiter, the line 
of apsides has shifted clockwise by more than 180\deg.
This allowed for dust measurements at varying Jovian local times.

The Jovian dust streams have been detected with DDS during all 33 Galileo 
orbits about Jupiter. Due to the DDS detection geometry,
dust stream particles could be mostly detected in the inner Jovian 
system, i.e. within $\sim 50\,\rm R_J$ from the planet, with a varying
detection geometry from orbit to orbit. Only after mid-2000
was the detection geometry significantly different so that dust stream 
particles could be detected during almost the entire Galileo orbit 
including apojove (cf. 
[\markcite{{\it Kr\"uger et al.}, 2003}]).

DDS was operated continuously during
Galileo's orbital tour, and highly time resolved real-time science 
(RTS) data were received from the instrument during most of the time 
periods when the geometry allowed for detection 
of the dust streams. Dust fluxes measured in the inner Jovian system
are shown in Fig. 1 
superimposed upon the Galileo
trajectory. In these time periods and spatial region the dust flux 
varied by more than four orders of magnitude 
($\sim 10^{-2}$ -- $10^{2} \, \rm m^{-2} \, sec^{-1}$). 
In order to calculate dust fluxes, the effective spin-averaged DDS 
sensor area 
[\markcite{{\it Gr\"un et al.}, 1992b, {\it Kr\"uger et al.}, 1999b}]
had to be taken into account, which did not remain constant as 
Galileo moved about Jupiter.
The time-dependent sensor area was 
calculated by taking the trajectory of a 10~nm particle as 
a 'standard` trajectory: particles of this size best explain the
observed impact directions of the streams 
[\markcite{{\it Gr\"un et al.}, 1998}].
When no dust flux is superimposed upon the Galileo trajectory 
in Fig. 1,
the dust streams were either not in the 
field of view of DDS or the spin-averaged effective sensor area 
was too small 
(below $20\, \rm cm^{2}$ which is less than 10\% of the maximum 
value) so that large uncertainties in the derived flux would occur.
No RTS dust data were collected during 
Galileo orbits 5 (Jan 1997) 
and 13 (Feb 1998) and a few short intervals of
spacecraft anomalies (safings).

\begin{figure}[t]
\vspace{-2cm}
\epsfxsize=0.96\hsize
\epsfbox{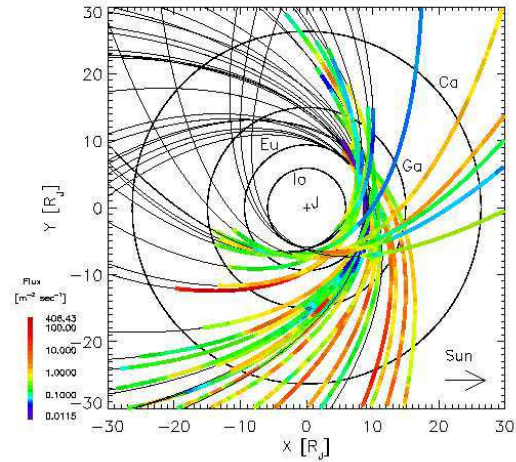}
\vspace{-2cm}
    \caption{
Galileo trajectory in the inner Jovian system from 1996 to
2002  
projected onto the planet's equatorial plane. 
The orbits of the Galilean moons are shown as circles. Dust fluxes measured with
DDS are color-coded superimposed upon the Galileo trajectory. 
A correction for instrument aging has 
been applied.
        }
    \label{orbitflux}
\end{figure}

The sensitivity of DDS has changed over time due to radiation-related 
aging effects in the instrument electronics. 
This instrument degradation has been revealed 
from an analysis of the electronics' response to simulated charge 
signals generated by a test pulse generator which is part of the DDS.
This change of the instrument response is consistent 
with a gradual shift of the charge signals measured for the Jovian 
dust streams during the Galileo Jupiter mission 
[\markcite{{\it Kr\"uger et al.}, in prep.}].
The drop in sensitivity
has been about a factor of 5 since Galileo's early Jupiter mission in 1996.
By assuming a size distribution of the particles, a correction
for the flux has been calculated and applied to the data:
taking the impact charge distribution measured for stream particles
in 1996 
([\markcite{{\it Kr\"uger et al.}, 2001}], their Fig.~6) when the
instrument still had its nominal sensitivity, the 
flux correction factor is one in 1996 and rises up to 25 in early 2002.

Figure 1
shows some systematic trends: some adjacent orbits
show similar dust fluxes at comparable spatial locations, but others 
do not. In order to search for systematic variations with Jovian local
time, we show the dust flux vs. local time for each orbit 
in Fig. 2.
Due to the orbital motion of Galileo, these 
data were collected at different distances from Jupiter as is indicated
by contour lines. For a source with a constant dust ejection rate 
and assuming continuity, one expects the dust flux to drop with the 
inverse square of the source distance. Therefore, in addition
to the corrections discussed above, the fluxes shown in 
Fig. 2
have been
multiplied by $r/(6\,\rm R_J)^2$ ($r$ being the distance from Jupiter) 
to eliminate the distance dependence of the flux due to the motion of 
Galileo. 

\begin{figure}
\epsfxsize=0.99\hsize
\epsfbox{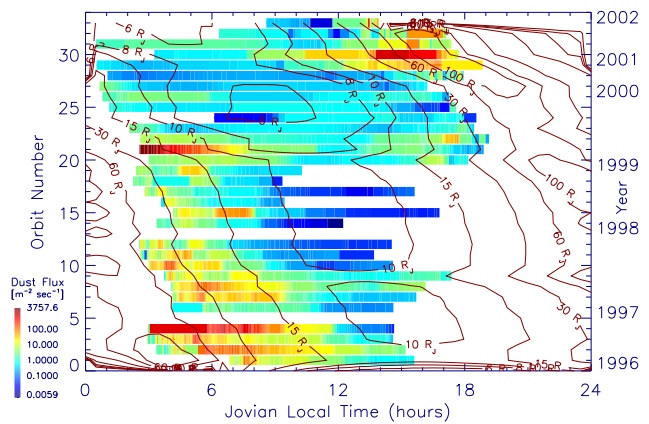}
    \caption{
Dust fluxes measured with DDS along 33 Galileo orbits about Jupiter vs. 
local time. Years are indicated on the right and  
spacecraft distance $r$ from Jupiter is indicated with red solid lines.
Fluxes were corrected for DDS aging and flux values
for less than $\rm 20\,cm^2$ 
sensor area were ignored.
A flux correction factor $(r/6\,\rm R_J)^2$ has been
applied.
}
    \label{fluxplot2}
\end{figure}

An additional cause for temporal flux variations is the dust
source Io itself: for example, during orbits 4 and 21, at distances 
between 15 and 30 $\rm R_J$, DDS measured the highest dust fluxes in 
the inner Jovian system during the entire Galileo mission, whereas
earlier and later adjacent orbits showed significantly lower fluxes. These 
variations are likely due to changes of the dust ejection of Io's 
volcanic plumes. However, a correlation of 
Io dust production with plume activity could not yet be 
established (see also 
[\markcite{{\it Kr\"uger et al.}, 2002}]). 
In spite of such orbit-to-orbit variability, 
systematic trends with Jovian local time are obvious in 
Fig. 2:
higher fluxes occurred on the dawn side of Jupiter whereas 
very low fluxes were typically measured around noon.

Flux variations from orbit to orbit caused by changes in the
Io dust production are expected to cancel out when 
averaging over the entire Galileo dust data set is performed.
This is done in Fig. 3:
the bottom panel shows
the dust flux for all Galileo orbits averaged over half-hour 
local time intervals. Before adding all flux measurements in a 
half-hour local time bin, the dust flux for each orbit was 
normalized to a range between 0 and 1. Due to this 
normalization, the flux correction for instrument aging 
does not have an effect on the averaged flux curve. Due to the 
changing detection geometry of DDS during the Galileo mission, the 
number of orbits available for
averaging varied with local time (cf. Fig. 2).
In order to avoid extreme flux values from individual orbits to 
appear in the average flux curve, an average data point is only 
shown when flux 
measurements are available from at least three Galileo orbits. 
Therefore, no data are 
shown between 20 and 24~h local time. In addition, flux measurements 
with an effective sensor area below $\rm 20\,cm^{2}$ were ignored.

The dust flux measurements from the entire Galileo mission presented in 
the bottom panel of Fig. 3
show a strong local 
time dependence: high fluxes occurred between 0 and 6~h and 
between  15 to 18~h. Systematically fewer particles 
were measured between 8 and 15~h. 

\begin{figure}[t]
\epsfxsize=0.71\hsize
\epsfbox{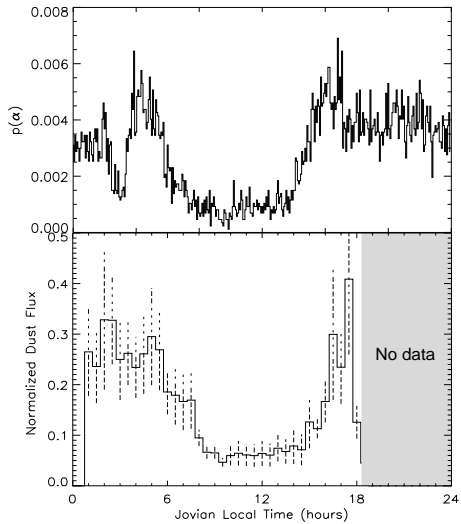}
    \caption{
Flux of Jovian dust streams as a function of Jovian local time. 
Top: distribution of dust particles crossing 
Ganymede's orbit within $\pm 3^{\circ}$ jovigraphic latitude 
derived from dynamical modelling [\markcite{{\it Hor{\'a}nyi et al.}, 1997}].
Bottom: Dust flux measurements averaged over up to 33 Galileo orbits.
A flux correction factor $(r/6\,\rm R_J)^2$ has been
applied.
}
    \label{fluxplot3}
\end{figure}

\section{Discussion}

We now compare the Galileo dust measurements with dynamical modeling 
[\markcite{{\it Hor{\'a}nyi et al.} 1993a, 1993b, 1997}].
Dust particles in the size range from 5 to 15~nm were released from
Io uncharged with the same initial velocity as Io's velocity at the
time of their ejection. Their charging and dynamics was followed 
numerically. The grains get charged in the Io plasma torus 
and their probability to escape the torus strongly depends on Io's 
position at the time of particle release 
[\markcite{{\it Hor{\'a}nyi et al.}, 1997}].
Jupiter's magnetic field has been described using the
GSFC~$\rm O_6$ or VIP~4 model 
[\markcite{{\it Connerney}, 1993}].
Jupiter's plasma has been  approximated using a plasma model, which is a
fit to the Voyager~1 and 2 plasma measurements
[\markcite{{\it Bagenal}, 1994}].

The local time dependence of the probability of the particles crossing 
Ganymede's orbit 
($r=\rm 15\,R_J$)  within $\pm 3^{\circ}$ above/below the equatorial 
plane of Jupiter  is shown in the top panel of
Fig. 3.
The data remarkably confirm the theoretical 
expectations: the local time dependence of the flux is
well in agreement, even the factor of $\sim 10$ contrast of the dust fluxes 
between 5 and 12~h local time is seen in the data. 


From the current study alone, we cannot distinguish between Io  as a 
point source or the entire plasma torus as a ring source. In fact, the long-term
averaging effectively smears Io dust into a ring. This study shows that
the plasma model, the description of the charging processes and the 
dynamics of the grains used in our model to describe dust production from
Io and its torus does reproduce the long-term averaged behavior of the
observations. 
In particular, the plasma conditions in
the Io torus are well represented as they are crucial for dust charging
there. Models without the torus' dawn-dusk
asymmetry do not 
show any  local time dependence of the dust flux.
Thus, dust stream particles can be used as tracers for the plasma 
conditions in the Io torus.
For example, we plan on using the direction and the magnitude of the 
dawn-to-dusk
displacement as free parameters to fit the simulation results to 
the observations.  
Hence, comparing the local time dependence of the measured flux to the 
theoretical
expectations allows for using dust measurements to independently confirm 
the direction and the magnitude of the cross-tail electric field.
In the same way the dust particles may also serve as 
probes of the Jovian magnetosphere. 

Our distance correction with $(r/\rm 6\,R_J)^2$ assumes an average source 
distance of Jupiter, consistent with a smeared out 'ring' source of the Io
torus. We also applied a correction factor $((r - \rm 6\,R_J)/\rm 6\,R_J)^2$
to the data, assuming a particle source at the Io torus location.
This does not significantly change the data in the bottom panel of 
Fig.~3. It even increases the contrast between 5 and 12~h local time.

We have also searched for systematic variations of particle size with
local time in the Galileo dust data set. Although the dust stream
particles are not in the calibrated mass and speed range of the 
dust instrument 
[\markcite{{\it Gr\"un et al.}, 1998}], the charge released upon 
impact of the dust grains onto the detector target can be used as an 
indicator for particle size. The data indicate a systematic variation 
of impact charge with local time, although much less pronounced than
in the case of the dust flux. Together with our simulations, 
this may also lead to better constraints of the plasma conditions in
the Io torus.

Here we have taken the entire DDS data set to average over flux 
variations from one orbit to the next in order to identify changes due to 
systematic variations alone.
The understanding of these systematic changes will allow us 
to `remove' them from the data and to identify  the `true' temporal 
variability of Io and the plasma torus.

Complementing our  earlier  time-series analysis,  our results also confirm
that Io and its plasma torus are the source for the majority (if not all) of 
the dust stream particles.
This source alone gives an excellent fit to the observed 
 spatial dependence of the dust fluxes, 
a significant contribution from an 
extended source, the gossamer ring, for example, or any other moon would show a different  ejection 
pattern.

A competing suggestion for the source of the stream particles was the
gossamer ring region of Jupiter. Here the plasma density is expected to
be very low, hence the dominant charging process is photoelectron
production from the grains. As the charging environment remains 
axially symmetric, grains -- if ejected from this region -- would
show a uniform ejection pattern.

In addition to the Galileo dust measurements within Jupiter's 
magnetosphere, the dust streams were also detected with three 
dust instruments in interplanetary space (Ulysses, Galileo and 
Cassini). The local time variation of the dust flux found in the
inner Jovian magnetosphere may at least partially explain
why Galileo (approaching Jupiter from $\rm \sim~6~h$ local time;
[\markcite{{\it Gr\"un et al.}, 1996}]) has measured on average 
higher fluxes than Ulysses 
(approaching from $\rm \sim 10~h$; 
[\markcite{{\it Gr\"un et al.}, 1993}]).
In 2004 Ulysses will 
approach Jupiter to 0.8~AU again and will be able
to measure the Jovian dust streams at varying 
local times and jovigraphic latitudes.
Although the interplanetary magnetic field affects the 
dynamics of the stream particles 
[\markcite{{\it Zook et al.}, 1996}],
the imprint of the 
local time variations may 
 show up that far from its origin. 

In summary, we now recognize the escaping dust particles as a  promising new 
diagnostic tool to characterize  the average  properties  of the Jovian
magnetosphere,  as well as the temporal variability of its most important 
source: Io.

%
%

\begin{acknowledgments}
We thank the Galileo project at JPL for effective and successful 
mission operations.
This research has been supported by the German Bundesministerium f\"ur Bildung 
und Forschung through Deutsches Zentrum f\"ur Luft- und Raumfahrt
(grant 50 QJ 9503 3). M.H. has been supported by NASA.
\end{acknowledgments}



\end{article}




\end{document}